\documentstyle[graphicx, 12pt]{article}
\begin{document}

\small
\hoffset=-1truecm
\voffset=-2truecm
\title{\bf The proximity force approximation for the Casimir energy
of plate-sphere and sphere-sphere systems in the presence of one
extra compactified universal dimension}
\author{Hongbo Cheng\footnote {E-mail address: hbcheng@ecust.edu.cn}\\
Department of Physics, East China University of Science and
Technology,\\ Shanghai 200237, China\\
The Shanghai Key Laboratory of Astrophysics, Shanghai 200234,
China}

\date{}
\maketitle

\begin{abstract}
The Casimir energies for plate-sphere system and sphere-sphere
system under PFA in the presence of one extra compactified
universal dimension are analyzed. We find that the Casimir energy
between a plate and a sphere in the case of sphere-based PFA is
divergent. The Casimir energy of plate-sphere system in the case
of plate-based PFA is finite and keeps negative. The
extra-dimension corrections to the Casimir energy will be more
manifest if the sphere is larger or farther away from the plate.
It is shown that the negative Casimir energy for two spheres is
also associated with the sizes of spheres and extra space. The
larger spheres and the longer distance between them make the
influence from the additional dimension stronger.
\end{abstract}

\vspace{3cm} \hspace{1cm} PACS number(s): 12.20.Ds.; 03.70.+k.;
11.10.Kk

\newpage

\noindent \textbf{I.\hspace{0.4cm}Introduction}

The existence of a very peculiar effect named as Casimir effect
was predicted in 1948. This effect shows the attraction between
two metallic uncharged parallel plates in vacuum. The existence of
such an attraction has been proved experimentally and the
measurements of the Casimir force have reached a higher accuracy
[2-9]. It should be pointed out that the simple device consisting
of parallel plates poses a certain degree of difficulty in the
experiment because of the difficulty in achieving parallelism
although a plenty of exact results for Casimir effect of parallel
plates were obtained theoretically. In fact the experimentalists
favour the configurations such as sphere-plate or sphere-sphere
system etc.. The straightforward computations of geometries
involving curved surfaces instead of parallel plates are
conceptually complicated. The proximity force approximation (PFA)
originated to give rise to a theoretical prediction for the
Casimir energy of the setup like sphere-plate or sphere-sphere
system as a function of the separation between the two components
becomes a standard tool for estimating the curvature effects for
nonplanar geometries in all experiments [10, 11]. Some powerful
methods have been developed to compute these Casimir interactions
further. These approaches include the semiclassical approximation
[12, 13], the optical path method [14-16], the world line approach
[17-20], the functional determinant or the multiple scattering
method [21-28] and the exact mode summation method [29, 30]. The
above issues are used to compute the corrections to the PFA. These
methods have been applied to derive the Casimir interactions of
the geometric configurations such as sphere-plane system [12-18,
20-22, 28], sphere-sphere configurations [13, 22, 24-27] etc. in
the four-dimensional spacetime.

In order to unify the interactions in nature, the Kaluza-Klein
model was put forward about eight decades ago [31,32]. This theory
introduced an additional compactified dimension to unify gravity
and classical electrodynamics. The string theory is also developed
to unify the quantum mechanics and gravity with the help of
introducing seven extra spatial dimensions. The gauge fields may
be localized on a four-dimensional brane, our real universe, and
only gravitons can propagate in the extra space transverse to the
brane [33, 34]. The existence and order of the additional
dimensions need to be explored both theoretically and
experimentally. The Casimir effect could open a window to observe
the extra space. The Casimir effects of the devices such as
parallel plates and cavities have been studied in the presence of
extra compactified dimensions and the influences from the extra
dimensions are shown [35-43].

It is fundamental to investigate the Casimir energies of the
sphere-plane and sphere-sphere systems in the world with
additional compactified dimension with the help of PFA. The known
Casimir energy could lead the further study of Casimir effect. As
the first step we choose the Kaluza-Klein model with one extra
dimension. We have to discuss the Casimir energy of system such as
sphere-plane or sphere-sphere ones in this model according to the
easily-established experiments although the theoretical
exploration is complicated. As the first step the PFA will be used
to estimate the Casimir energy between one sphere and a plate or
two spheres in the five-dimensional spacetime and the corrections
from the additional compactified dimension must appear. The PFA
should be adequate because the technique gives rise to the leading
terms in the derivative expansion of the Casimir energy of
nonplanar geometries. The PFA results can afford the acceptable
description of the properties. Further the PFA results can be
improved. We open a window to probe the extra dimensions in the
Casimir effect for nonplanar system except for parallel plates.
The purpose of this paper is to reexamine the PFA for the Casimir
energy of sphere-plate and sphere-sphere systems in the universe
involving the fifth compactified spatial dimension. We wonder how
the fifth dimension modifies the Casimir energy. We sum up the
Casimir energies for a serious of parallel plates to obtain the
Casimir energy of the system under PFA. In particular we can
wonder which method of plate-based PFA or sphere-based PFA is
adequate for the sphere-plate case in the Kaluza-Klein model. We
also compare the Casimir energy in the presence of one additional
compactified with that in the four-dimensional spacetime in order
to exhibit the extra-dimension corrections theoretically. These
energies belong to the kinds of system consisting of plate and
sphere or two spheres respectively. The discussions and
conclusions are emphasized in the end.

\vspace{0.8cm} \noindent \textbf{II.\hspace{0.4cm}The PFA for the
Casimir energy of plate-sphere system in the presence of one extra
compactified universal dimension}

Within the frame of Kaluza-Klein approach the scalar field
satisfying the Dirichlet boundary conditions on the parallel
plates was studied in the spacetime with only one extra dimension,
the Casimir energy density is [35, 37-39],

\begin{equation}
\varepsilon_{C}(R)=-\frac{\pi^{2}}{720}\frac{1}{R^{3}}
+\frac{1}{16\pi^{5}}\Gamma(2)\zeta(4)\frac{1}{L^{3}}
-\frac{1}{16\pi^{\frac{13}{2}}}\Gamma(\frac{5}{2})\zeta(5)\frac{R}{L^{4}}
-\frac{1}{4\pi^{2}}\frac{1}{RL^{2}}\sum_{n_{1},n_{2}=1}^{\infty}
(\frac{n_{2}}{n_{1}})^{2}K_{2}(2\frac{R}{L}n_{1}n_{2})
\end{equation}

\noindent where $R$ is the separation of the two parallel plates
and $L$ is the radius of the extra dimension. $K_{\nu}(z)$ is the
modified Bessel functions of the second kind and falls
exponentially with $z$. This Casimir energy density can be
inserted into the surface integral to obtain the Casimir energy of
two arbitrary smooth surfaces under PFA like [9, 22],

\begin{equation}
\varepsilon_{PFA}=\int\int_{A}\varepsilon_{C}[z(\sigma)]d\sigma
\end{equation}

\noindent where the scalar field also obey the Dirichlet boundary
conditions on the smooth surfaces. $A$  stands for the area of one
of the opposing surfaces which are locally separated by the
distance $z(\sigma)$. $\varepsilon_{C}[z(\sigma)]$ represents the
corresponding Casimir energy density and is also
surface-dependent. It should be pointed out that the local
distance vector $\vec{z}(\sigma)$ is perpendicular only to the
plate segment $d\sigma$ which is tangential to only one of the
surfaces.

In the case of two surfaces with different shapes, the Casimir
energy $\varepsilon_{PFA}$ will not be uniquely defined. For
example the surfaces in the sphere-plate system are different, so
there will be two kinds of expressions of Casimir energy for the
sphere-based PFA and plate-based PFA respectively. In the case of
sphere-based PFA, the Casimir energy expression is [22],

\begin{equation}
E_{S0}=-\frac{\pi^{3}}{720}\frac{a}{d^{2}}
\{1-\frac{3d}{a}-6(\frac{d}{a})^{2}[1-(1+\frac{d}{a})
\ln(1+\frac{a}{d})]\}
\end{equation}

\noindent On the other hand, the Casimir energy for the
plate-based PFA was expressed as [22],

\begin{equation}
E_{P0}=-\frac{\pi^{3}}{720}\frac{a}{d^{2}}\frac{1}{1+\frac{d}{a}}
\end{equation}

\noindent where $a$ is the radius of sphere and $d$ is the
shortest distance between the sphere and the plate.

We start to discuss the extra-dimension corrections to the Casimir
energy of plate-sphere system with PFA. We let the Casimir energy
density in Eq. (1) replace the integrand in Eq. (2). In the case
of sphere-based PFA, the Casimir energy is,

\begin{eqnarray}
E_{S}=\int\int_{half-sphere}\varepsilon_{C}(\vec{z})
a^{2}d\Omega\hspace{5cm}\nonumber\\
=-\frac{\pi^{3}}{720}\frac{\xi}{\mu^{2}}\{1-\frac{3\mu}{\xi}
-6(\frac{\mu}{\xi})^{2}[1-(1+\frac{\mu}{\xi})\ln(1+\frac{\xi}{\mu})]\}
\frac{1}{L}\hspace{1.5cm}\nonumber\\
+\frac{1}{8\pi^{4}}\Gamma(2)\zeta(4)\xi^{2}\frac{1}{L}\hspace{7cm}\nonumber\\
-\frac{1}{2\pi}\frac{\xi^{2}}{L}\sum_{n_{1},n_{2}=1}^{\infty}
(\frac{n_{2}}{n_{1}})^{2}\int_{0}^{1}\frac{x}{\mu+\xi-\xi x}
K_{2}(2n_{1}n_{2}\frac{\mu+\xi-\xi x}{x})dx\nonumber\\
+\frac{1}{8\pi^{\frac{11}{2}}}\Gamma(\frac{5}{2})\zeta(5)
\xi^{2}[\int_{0}^{\frac{\pi}{2}}(\xi-\frac{\mu+\xi}{\cos\theta})
\sin\theta d\theta]\frac{1}{L}\hspace{2cm}
\end{eqnarray}

\noindent with the choice of $|\vec{z}|=\frac{d+a}{\cos\theta}-a$
and here

\begin{equation}
\xi=\frac{a}{L}
\end{equation}

\noindent and

\begin{equation}
\mu=\frac{d}{L}
\end{equation}

\noindent It is clear that the last term in the equation above,

\begin{equation}
\frac{1}{8\pi^{\frac{11}{2}}}\Gamma(\frac{5}{2})\zeta(5)
\xi^{2}[\int_{0}^{\frac{\pi}{2}}(\xi-\frac{\mu+\xi}{\cos\theta})
\sin\theta d\theta]\frac{1}{L}=\infty
\end{equation}

\noindent is divergent. The Casimir energy of plate-sphere system
under the sphere-based PFA certainly approaches the infinity
because of the divergent term. We introduce the plate-based PFA to
this system, then the Casimir energy becomes,

\begin{eqnarray}
E_{P}=\int\int_{x^{2}+y^{2}\leq a^{2}}\varepsilon_{C}(\vec{z})
dxdy\hspace{4cm}\nonumber\\
=-\frac{\pi^{3}}{720}\frac{\xi}{\mu^{2}}\frac{1}{1+\frac{\mu}{\xi}}
\frac{1}{L}\hspace{6cm}\nonumber\\
+\frac{1}{16\pi^{4}}\Gamma(2)\zeta(4)\xi^{2}\frac{1}{L}
-\frac{1}{8\pi^{\frac{11}{2}}}\Gamma(\frac{5}{2})\zeta(5)
(\frac{1}{2}\xi^{2}\mu+\frac{1}{6}\xi^{3})\frac{1}{L}\nonumber\\
-\frac{1}{2\pi}\frac{1}{L}\sum_{n_{1},n_{2}=1}^{\infty}
(\frac{n_{2}}{n_{1}})^{2}\int_{\mu}^{\mu+\xi}\frac{\mu+\xi-t}{t}
K_{2}(2n_{1}n_{2}t)dt
\end{eqnarray}

\noindent while we choose
$|\vec{z}|=d+a-\sqrt{a^{2}-x^{2}-y^{2}}$. This Casimir energy is
finite. If the plate and the sphere are sufficiently far away from
each other, the asymptotic behaviour of the casimir energy is,

\begin{eqnarray}
\lim_{\mu\longrightarrow\infty}E_{P}=-\frac{1}{16\pi^{\frac{11}{2}}}
\Gamma(\frac{5}{2})\zeta(5)\xi^{2}\mu\frac{1}{L}\nonumber\\
<0\hspace{5cm}
\end{eqnarray}

\noindent The Casimir energy of plate-sphere device with
plate-based PFA in the spacetime with one extra dimension is
depicted in Fig. 1. The curves of energy are similar with
different size of sphere. The larger spheres make the Casimir
energy higher. The nature of the Casimir energy keeps negative no
matter how far the plate and the sphere localize each other. When
the two components are moved farther, the value of the Casimir
energy is linearly decreasing like in Eq. (10). In order to show
the influence from the additional dimension evidently, we compare
the two kinds of energies with or without extra space. We
demonstrate the ratio $h=\frac{E_{P}}{E_{P0}}$ in Fig. 2. It is
interesting that the larger sphere and the larger distance between
the plate and sphere both make the ratio larger, and the larger
ratio exhibits the extra-dimension correction more manifest.

\vspace{0.8cm} \noindent \textbf{III.\hspace{0.4cm}The PFA for the
Casimir energy of sphere-sphere system in the presence of one
extra compactified universal dimension}

Now we pay our attention to the Casimir energy between two spheres
under PFA in the world involving one additional compactified
dimension. At first we write the Casimir energy of sphere-sphere
system under PFA in the four-dimensional spacetime as follow
according to Ref. [22],

\begin{equation}
E_{SS0}=-\frac{\pi^{3}}{720}\frac{a^{2}}{d^{2}(2a+d)}
\end{equation}

\noindent It is necessary that we take the CAsimir energy density
for parallel plates subject to one extra dimension in Eq. (1) the
place of the integrand in Eq. (2). We derive the Casimir energy of
the sphere-sphere system as,

\begin{eqnarray}
E_{SS}=\int\int_{x^{2}+y^{2}\leq
a^{2}}\varepsilon_{C}[2z(\sigma)]d\sigma\hspace{4cm}\nonumber\\
=-\frac{\pi^{3}}{720}\frac{\xi^{2}}{\mu^{2}(2\xi+\mu)}\frac{1}{L}
\hspace{6cm}\nonumber\\
+\frac{1}{16\pi^{4}}\Gamma(2)\zeta(4)\xi^{2}\frac{1}{L}
-\frac{1}{8\pi^{\frac{11}{2}}}\Gamma(\frac{5}{2})\zeta(5)
(\frac{1}{2}\xi^{2}\mu+\frac{1}{3}\xi^{3})\frac{1}{L}\nonumber\\
-\frac{1}{4\pi}\frac{1}{L}\sum_{n_{1},n_{2}=1}^{\infty}(\frac{n_{2}}{n_{1}})^{2}
\int_{\mu}^{\mu+\xi}\frac{\mu+\xi-t}{t}K_{2}(4n_{1}n_{2})dt
\end{eqnarray}

\noindent where $\sigma=\sigma(x, y)$ and $d\sigma=dxdy$. Here the
local distance is denoted as
$\vec{z}(x,y)=\frac{d}{2}+a-\sqrt{a^{2}-x^{2}-y^{2}}$. $a$ is a
common radius of the two spheres and $d$ is the shortest
separation between the sphere surfaces. We can move the two
spheres farther away from each other to investigate the asymptotic
value of the Casimir energy as follow,

\begin{eqnarray}
\lim_{\mu\longrightarrow\infty}E_{SS}=-\frac{1}{16\pi^{\frac{11}{2}}}
\Gamma(\frac{5}{2})\zeta(5)\xi^{2}\mu\frac{1}{L}\nonumber\\
<0\hspace{5cm}
\end{eqnarray}

\noindent which is the same as the results of plate-sphere system
under the plate-based PFA in Eq. (10). We show the Casimir energy
between two spheres controlled by PFA in the presence of one
additional dimension graphically in Fig. 3. The shapes of the
energy associated with the size of the spheres are similar. The
Casimir energy becomes higher as the two spheres enlarge. The sign
of the Casimir energy of the two spheres remains negative. The
Casimir energy will also be linearly decreasing as in Eq. (13) if
the distance between the spheres is longer. We can also make use
of the sphere-sphere experiment to explore the extra space. We
plot the ratio $h=\frac{E_{SS}}{E_{SS0}}$ in Fig. 4 to compare the
two-sphere Casimir energy including or excluding the fifth
dimension. It is found that the larger ration is due to the larger
spheres which localize farther each other. The larger ratio
originated by the extra dimensions could be measured in the
experiment.

\vspace{0.8cm} \noindent \textbf{IV.\hspace{0.4cm}Discussion and
conclusion}

In this work it is the first time to investigate the Casimir
energies of plate-sphere and sphere-sphere under PFA in the
presence of one extra compactified universal dimension. Having
studied the Casimir energy of plate-sphere device in the cases of
sphere-based PFA and plate-based PFA respectively, we discover
that the Casimir energy belonging to the sphere-based PFA is
divergent, but the energy for plate-based PFA is finite and keeps
negative. It is shown that the Casimir energy of plate-sphere
system governed by plate-based PFA depends on the experimental
structure and the additional dimension. Further we find that the
larger sphere and longer distance between plate and sphere make
the influence from extra dimension more obvious. We also
scrutinize the Casimir energy between two identical spheres
limited by PFA in the spacetime with one extra compactrified
dimension. We demonstrate that the negative Casimir energy has
something to do with the radii of the spheres and the gap between
them. It is similar to the results of the plate-sphere system in
the case of plate-based PFA that the extra-dimension corrections
will become more visible in the experiment including larger
spheres which are farther away from each other. Our predictions
from PFA are leading terms of Casimir energies and certainly can
become a window to explore the extra compactified space. The
related topics need further research and in progress.

\vspace{1cm}
\noindent \textbf{Acknowledge}

The author thanks Professor K. Milton and Professor M. Bordag for
helpful discussions. This work is supported by NSFC No. 10875043
and is partly supported by the Shanghai Research Foundation No.
07dz22020.

\newpage
\begin{figure}
\setlength{\belowcaptionskip}{10pt} \centering
\includegraphics[width=15cm]{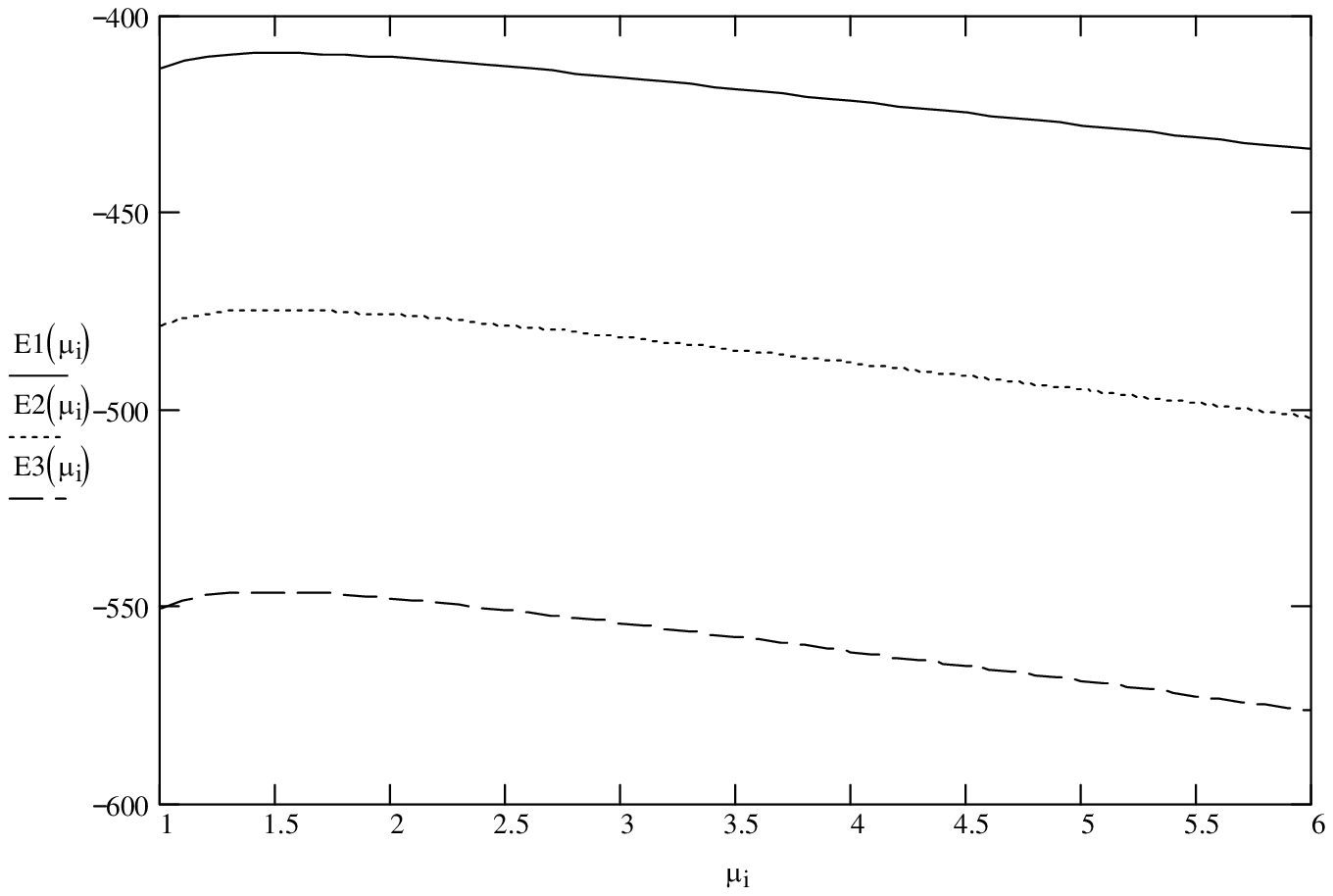}
\caption{The solid, dot and dashed curves of the Casimir energy of
plate-sphere system in unit of $\frac{1}{L}$ as functions of ratio
of the shortest distance between the plate and sphere and
extra-dimension radius $\mu=\frac{d}{L}$ under the plate-based PFA
for $\xi=200, 210, 220$ respectively.}
\end{figure}

\newpage
\begin{figure}
\setlength{\belowcaptionskip}{10pt} \centering
\includegraphics[width=15cm]{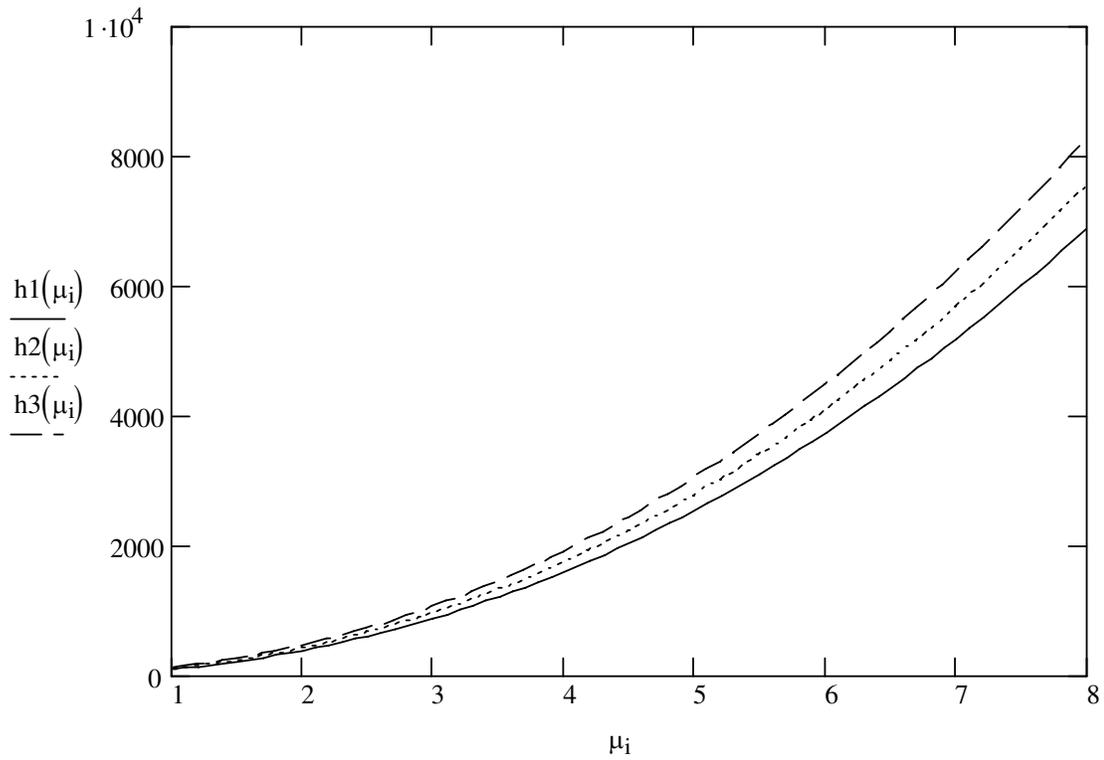}
\caption{The solid, dot and dashed curves of the ratio of Casimir
energies of plate-sphere system under the plate-based PFA with and
without the additional dimension as functions of $\mu=\frac{d}{L}$
for $\xi=200, 210, 220$ respectively.}
\end{figure}

\newpage
\begin{figure}
\setlength{\belowcaptionskip}{10pt} \centering
  \includegraphics[width=15cm]{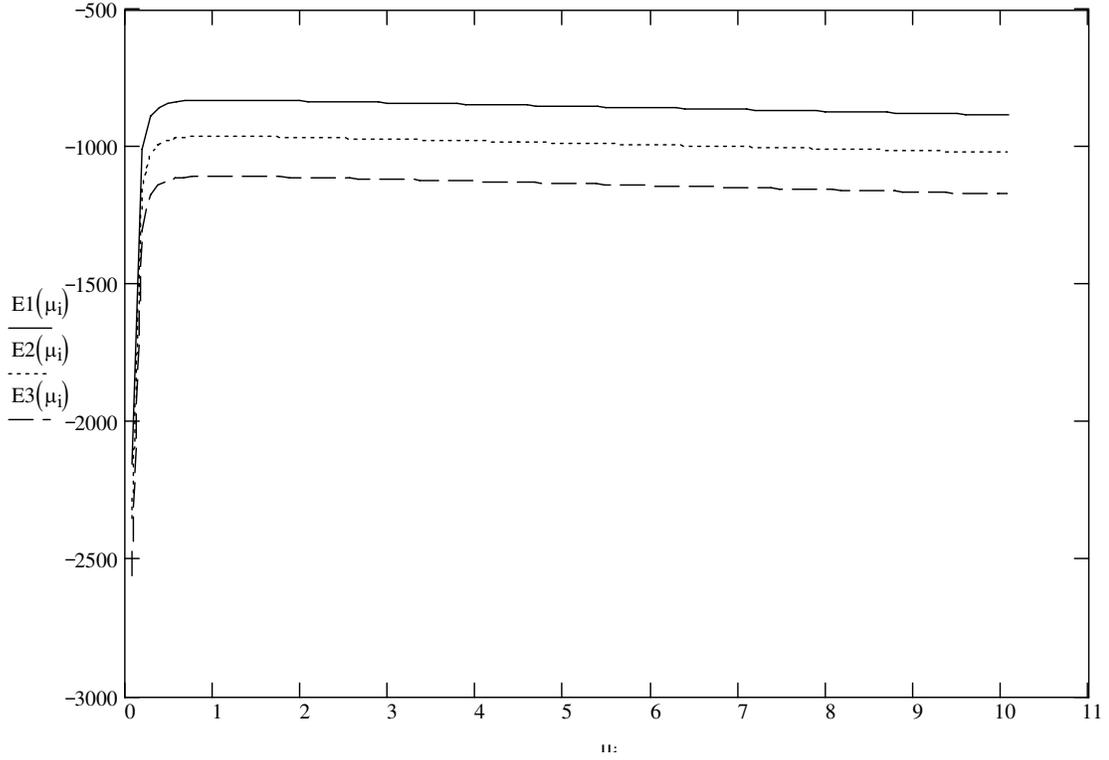}
  \caption{The solid, dot and dashed curves of the Casimir energy of
two spheres system limited by PFA in unit of $\frac{1}{L}$ as
functions of ratio of the shortest distance between two spheres
and extra-dimension radius $\mu=\frac{d}{L}$ under the plate-based
PFA for $\xi=200, 210, 220$ respectively.}
\end{figure}

\newpage
\begin{figure}
\setlength{\belowcaptionskip}{10pt} \centering
  \includegraphics[width=15cm]{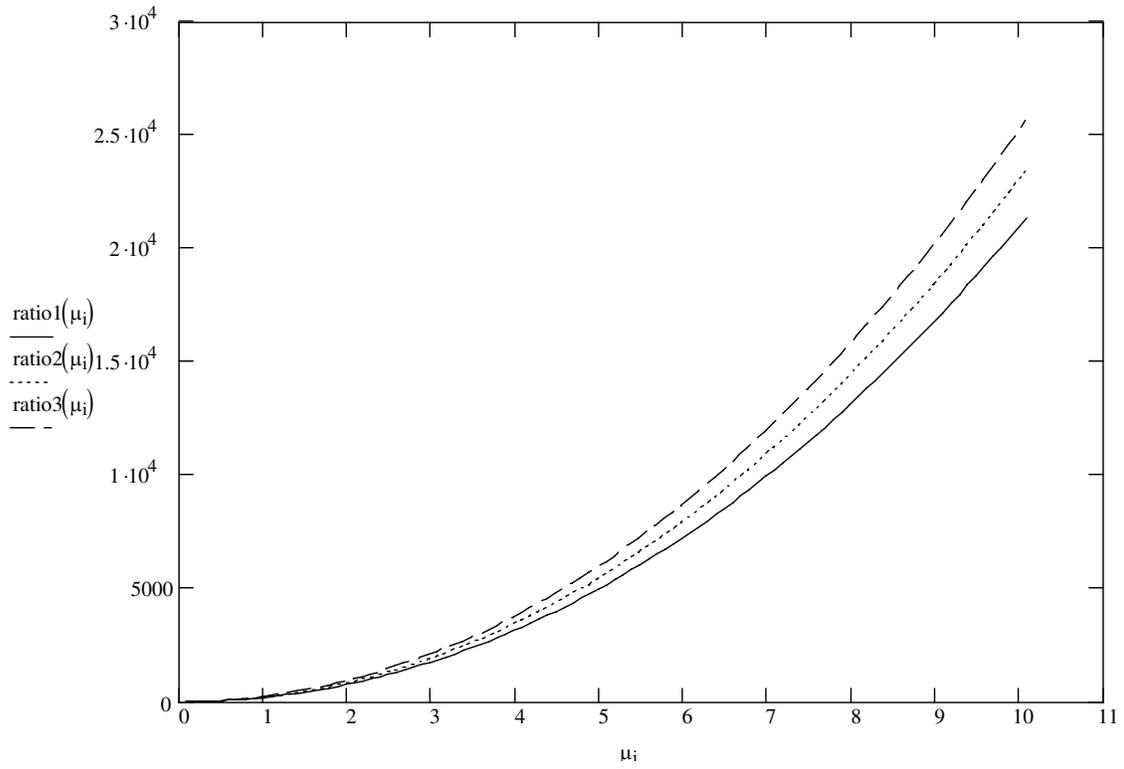}
  \caption{The solid, dot and dashed curves of the ratio of Casimir
energies of two spheres system limited by PFA with and without the
additional dimension as functions of $\mu=\frac{d}{L}$ for
$\xi=200, 210, 220$ respectively.}
\end{figure}

\end{document}